\documentclass{article}%
\usepackage{graphicx}
\usepackage[dvips]{color}

\begin{document}

\title{Galactic mapping with general relativity and the observed rotation curves}

\author{N. S. Magalhaes
\\Department of Exact and Earth Sciences
\\ Federal University of Sao Paulo, Diadema, SP, Brazil
\\and
\\F. I. Cooperstock
\\Department of Physics and Astronomy
\\ University of Victoria, Victoria, BC, Canada}

\date{July 31, 2015}

\maketitle

\begin{abstract}
Typically, stars in galaxies have higher velocities than predicted by Newtonian gravity in conjunction with observable galactic matter. To account for the phenomenon, some researchers modified Newtonian gravitation; others introduced dark matter in the context of Newtonian gravity. We employed general relativity successfully to describe the galactic velocity profiles of four galaxies: NGC 2403, NGC 2903, NGC 5055 and the Milky Way. Here we map the density contours of the galaxies, achieving good concordance with observational data. In our Solar neighbourhood, we found a mass density and density fall-off fitting observational data satisfactorily.  From our GR results, using the threshold density related to the observed optical zone of a galaxy, we had found that the Milky Way was indicated  to be considerably larger than had been believed to be the case.  To our knowledge, this was the only such existing theoretical prediction ever presented. Very recent observational results by Xu et al. have confirmed our prediction. As in our previous studies, galactic masses are consistently seen to be higher than the baryonic mass determined from observations but still notably lower than those deduced from the approaches relying upon dark matter in a Newtonian context. In this work, we calculate the non-luminous fraction of matter for our sample of galaxies that is derived from applying general relativity to the dynamics of the galaxies.   The evidence points to general relativity playing a key role in the explanation of the stars' high velocities in galaxies. Mapping galactic density contours directly from the dynamics opens a new window for predicting galactic structure.
\end{abstract}

\section{Introduction}

The discovery of the existence of remarkably high stellar velocities was achieved from the analysis of galactic rotation curves \cite{sofrub01}, plots of stellar velocities as a function of their distances from the galactic axis of rotation. Theoretical approaches were devised in the investigation of this mystery, including MOND (see for e.g. \cite{fam12}) and the proposal of dark matter (see for e.g. \cite{ber}). The first introduces an alternative theory to that of Newtonian gravity (NG), while the second retains NG proposing the existence of a new kind of matter, designated as dark matter, to provide the extra driving force to induce the high velocities. 

Another approach \cite{ct07} applies general relativity (GR) to fit the rotation curves. It is 
the general consensus of the great majority of the physics community that general relativity is our best theory of gravity. However, since in galaxies the gravitational fields are generally weak and their constituent stellar velocities, while very high by our terrestrial standards, are not relativistic, it is normally assumed that NG should suffice to describe galactic gravitational phenomena. While the constraints involving weak-fields and non-relativistic-velocities generally suffice for NG application, it has been shown \cite{ct07} that for freely gravitating extended systems of matter under rotation, the dynamical predictions arising from GR depart  significantly from those employing Newtonian gravity. Our methodology has been vetted by others (see, for e.g. \cite{bg}).

Using GR, it was possible to fit four rotation curves \cite{ct07} and, subsequently, the approach yielded good fits to three additional galaxies \cite{cc12}, with all but one of the subsequent critiques addressed\footnote{Apart from a paper by Fuchs and Phleps \cite{fp} whose critique (addressed here) was well-founded, in a very recent paper \cite{row}, it is claimed that we and others were wrong and that NG suffices to analyze galactic rotation curves.  In a forthcoming paper, we show that this author is mistaken.}. From these fits, the deduced masses were consistently lower than those proposals requiring large stores of dark matter.

In this paper we go beyond the previous works and utilize the now widely expanded store of observational rotational velocity data to probe galactic structures in greater detail while fitting the rotation curves of four galaxies using GR alone. We compare our results with observational data of stellar concentrations and the concordance fortifies our confidence in the correctness of the procedure. As in our previous studies, the galactic  masses that we deduce continue to be less than those presented using NG. As a new feature in this paper, we highlight the mapping of the density as providing a means of determining unseen structures.

The enhanced store of data has enabled us to further apply the criterion of optical galactic radius determination at the threshold density of $10^{-21.75}$ kg m$^{-3} = $ 0.0026 M$_{\odot}$ pc$^{-3}$. This number is close to the average density of the interstellar matter in the vicinity of the Sun in the densest molecular regions, where the density varies from\cite{fer} $\sim 10^{-20.7}$ kg m$^{-3} = $ 0.029 M$_{\odot}$ pc$^{-3}$ to $\sim 10^{-22.7}$ kg m$^{-3} = $ 0.00029 M$_{\odot}$ pc$^{-3}$. We suggest that the presence of this number in the model may be physically tied to the brightness of molecular clouds in the galaxies. Applying this criterion to the Milky Way, we found that our own Galaxy was indicated to be larger than is generally believed, in fact consistent with the recently announced observations of Xu et al. \cite{xu} The prediction of the larger  MW fortifies our confidence in GR as the premier theory of gravity. 
					
Besides producing well-fitted rotation curves, we found that the mass density at a radius of $\sim 7.7$ kpc from our Galaxy's nucleus is $5.8 \times 10^{-21}$ kg m$^{-3} =$  0.085 M$_{\odot}$ pc$^{-3}$. This value is close to the observed one\cite{hf} of $6.3 \times 10^{-21}$ kg m$^{-3} = $ 0.094 M$_{\odot}$ pc$^{-3}$ where our solar position is believed to lie, between $7.3$ and $8.5$ kpc from the nucleus. Also, the density fall-off in the direction perpendicular to the Galactic plane has good concordance with the observational data near our sun within the limits of applicability of the model. 

Regarding the mass-to-light ratio (M/L), assuming that it equals 1, we found that all galaxies in the sample are expected to present non-luminous matter in variable proportions. This detailed quantitative result was absent in \cite{ct07} and \cite{cc12}, although in those papers the role of dark matter within the rotation curve problem was questioned. We found that according to our GR model, the amount of non-luminous matter in the galaxies is notably smaller than the amount of dark matter expected in the context of Newtonian gravity. As we argue in the text, this is a natural consequence of the fact that general relativistic effects are significant in the physics that yields rotation curves. 
	
The plan of the paper is as follows: in Section \ref{sec:intro}, we introduce the primary general relativistic equations for developing the rotation curves and the underlying galactic density distributions that are used in the following studies. The details leading up to these equations and the rationale for invoking a stationary axially-symmetric model of a rotating pressure-free fluid as a basis to model a galaxy are dealt with in our earlier papers \cite{ct07,cc12} and is summarized in Appendix \ref{ap:1}.

In Sections \ref{sec:2403}, \ref{sec:2903} and \ref{sec:5055}, we respectively analyze galaxies NGC 2403, NGC 2903 and NGC 5055. Each galaxy has its own particular characteristics that were useful to display.

In Section \ref{sec:mw}, we turn to the Milky Way, which has its particular challenges stemming from our inability to view it from outside. However, we have more detailed density data from our advantage in being within. This has enabled us to test our model. We conclude with a discussion in Section \ref{sec:con}.

\section{The galactic model}\label{sec:intro}

In our approach to the problem, we assume that a galaxy's components behave and interact like a low density, freely gravitating fluid without pressure, rotating in stationary, axially-symmetric motion. This represents an idealized approach to the actual detailed complex galactic dynamics that prevails but it allows a general relativistic solution. The use of axially-symmetric systems of stars in rotation to study galactic disks is not new \cite{at} and it provides a powerful tool for a lowest-order approximation. 

	We present technical details on the approach in appendix \ref{ap:1}. The metric function relevant for the rotational motion, $N(r,z)$, can be related to a star's tangential velocity, $V$,  according to
\begin{equation}
	N	= \frac{rV}{c}
	\label{eq1}
\end{equation}
in the presence of weak fields. The cylindrical polar coordinates are $(r, \phi , z)$ and $c$ is the speed of light in vacuum.

	A functional form can be found for $N$ by solving one of Einstein's equations, yielding the following expression for $V(r, z)$:
\begin{equation}
	V= -c\sum_{n} k_n C_n e^{-k_n |z|}J_1(k_nr),
\label{eq2}
\end{equation}
where the $J_1$ are the first-order Bessel functions and $k_n$ are the zeroes of $J_0$, the zeroth-order Bessel functions at the $r$ limits of integration. This is the basic equation that we used to obtain our results. By applying actual data points $(r,V)$ to equation (\ref{eq2}) assuming $z = 0$ (invoking the standard approximation of positioning the stars in the galactic midplane), we found the corresponding $C_n$-coefficients through the use of the \verb"bestfit" function of Maple\footnote{Most of the computations in this paper were performed using maple(TM), release 11.01. Maple is a trademark of Waterloo Maple Inc. Maplesoft, a division of Waterloo Maple Inc., Waterloo, Ontario.}. These coefficients were substituted back into equation (\ref{eq2}) up to a chosen number of parameters ($n$) and, using equation (\ref{eq1}), the resulting equation yielded an approximate expression for $N(r,z)$.

A galaxy's mass density function was calculated from equation (\ref{eq:a5}),which can be rewritten as 
\begin{equation}
\rho(r, z) = \frac{c^2}{8 \pi G} \frac{N_r^2 + N_z^2}{r^2}, 
\label{eq3}
\end{equation}
by substituting the partial derivatives ($N_r$ and $N_z$) of the approximated expression obtained for $N(r, z)$, with $G$ being Newton's gravitational constant. From equation (\ref{eq3}) we calculated, by volume integration, the galaxy's total mass.

	We also used the obtained density functions to check a hypothesis introduced and tested in previous works \cite{ct07, cc12}, where it was noted that a certain value of the mass density appeared to be related to the optical luminosity threshold for galaxies as tracked in the radial direction. In other words, for six galaxies investigated thus-far, the mass density was found to be $\rho _{opt} \sim 10^{-21.75}$ kg m$^{-3}$ = 0.0026 M$_\odot$ pc$^{-3}$ at $z = 0$  and near the radius where the surface brightness in the {\it r-band} of the galaxies reached a minimum value.
	
	In the present work we use the definition of the diameter of a disc galaxy given by the value of the logarithm of the length of the projected major axis of the galaxy at the isophotal limiting surface brightness of $25$ mag arcsec$^{-2}$ \cite{pa}. This definition reduces all visual diameters to a common standard system. Applying cataloged information to this definition, we calculated the optical radius of a galaxy from observed quantities, as we show in appendix \ref{ap:2}. This value was used in comparison with the value of the radius, $r_{opt}$, deduced from the position  where $\rho _{opt}$ occurred in our model, thus providing new tests to the hypothesis. In units of solar luminosities per squared parsec, in the {\it r-band} one finds $25$ mag arcsec$^{-2} =$ 2.59 L$_\odot$ pc$^{-2}$.
	
As pointed out earlier, the density $10^{-21.75}$ kg m$^{-3}$ is close to the average density of the interstellar matter in the vicinity of the Sun in the densest molecular regions, where this density varies from $\sim 10^{-20.7}$ to $\sim 10^{-22.7}$ kg m$^{-3}$. We, therefore, suggest that the presence of this number in the model may be physically tied to the brightness of molecular clouds in the galaxies, a prediction awaiting more observational evidence.

As to be expected, and as we witnessed in our earlier works, our general relativistic model predicts more mass than that which is predicted on the basis of the visible stellar matter alone. An excess in the form of non-luminous matter (NLM) has been referred to as dark matter (DM) by others who have estimated its extent on the basis of NG. Moreover, because of the very large extent of the DM component predicted by NG, DM has come to be regarded as solely gravitationally interacting (and at most possibly also weakly interacting) matter outside of our known standard model lexicon. 
\footnote{
The currently designated ``dark matter'' was originally referred to by researchers as ``exotic dark matter'' but the ``exotic'' adjective has been dropped. This is unfortunate as there is a need to refer to normal baryonic matter that is non-luminous, i.e. ``dark''. For this reason, we have found it necessary to introduce the descriptor ``NLM'' for this normal non-luminous matter to fill the created void.
} 
On the other hand, with our more modest galactic mass measures derived on the basis of general relativistic dynamics, one might consider the more conservative suggestion that the excess NLM could be undetected normal baryonic matter after all. We adopt the NLM designation as opposed to the usual DM one to underline our maintaining an open mind as to the true nature of this extra matter.

\section{NGC 2403}\label{sec:2403}

The rotation curve (RC) of NGC 2403 was studied with other approaches \cite{db,bo} and has a substantial number of data points available \cite{db,fr}. All the data sets which we used are listed in appendix \ref{ap:3}. We also use this galaxy as an example and the procedures we used with it, detailed below, were applied to the other galaxies which we will present.

One essential issue to address was the determination of the ideal number of parameters ($n$) to limit the expansion in equation (\ref{eq2}) and best-fit the RC. After checking the curve fit with a number of parameters, we determined the smallest $n$ that showed a good visual fit to the data: $n=13$. The coefficients that we used for our sample of galaxies, obtained with Maple, are listed in appendix \ref{ap:4}.

\begin{figure}
\begin{center}
\includegraphics[width=4.5in]{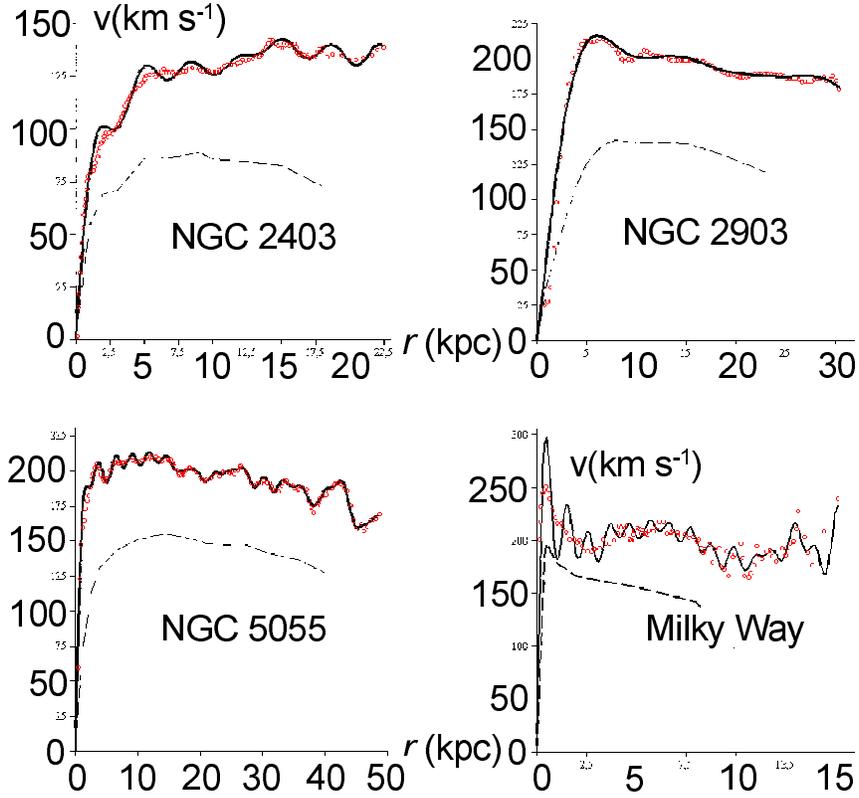}
\end{center}
\caption{
\label{fig:RCs}
	The rotation curve fits (solid lines) are our results, obtained from the expansion of the GR velocity functions for the galaxies NGC 2403, NGC 2903, NGC 5055 and the Milky Way. The circles are actual data points. The dashed line is the rotation curve that is obtained by applying the GR mass density to Newtonian gravitational theory.
	}
\end{figure}

\begin{figure}
\begin{center}
\includegraphics[width=4.5in]{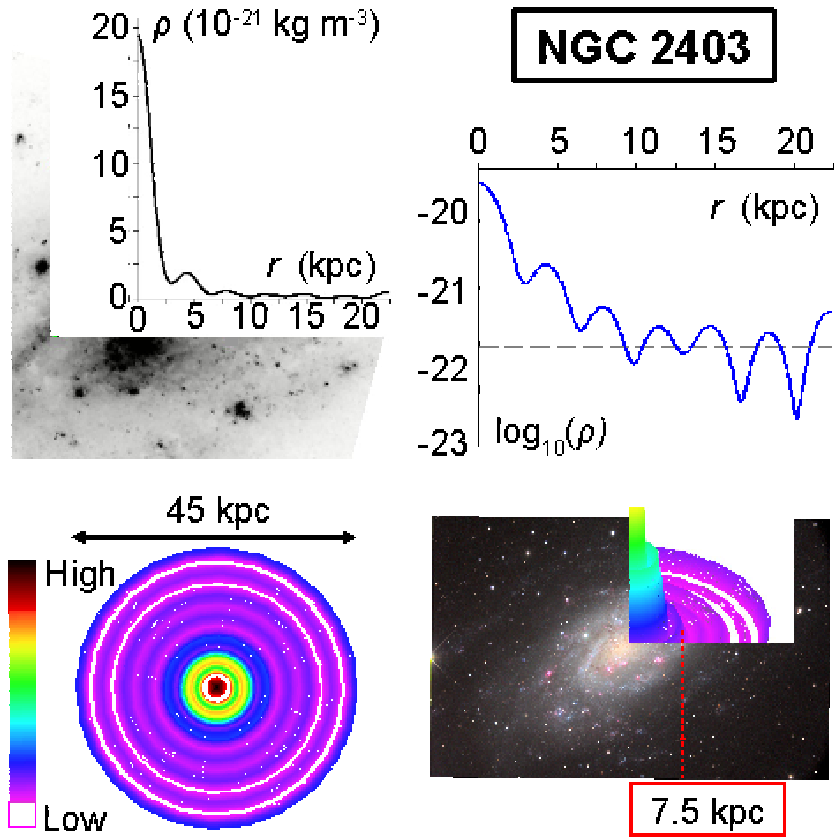}
\end{center}
\caption{
\label{fig:results2403}
Results for the galaxy NGC 2403. The plot in the top left figure shows the volume density curve at $z=0$ derived from the expansion of the velocity function and an image of the galaxy in the infrared band is included in the figure for comparison. The top right figure presents the logarithmic plot (solid line) for the mass density at $z = 0$, with the dashed line located at -21.75. The contour plot in the bottom left figure refers to the density distribution in the midplane, with the density varying between $\sim 2.2 \times 10^{-23}$ (white) and  $\sim 2 \times 10^{-20}$ kg m$^{-3}$ (black). A combination of the contour plot (rotated) with a picture of the galaxy is presented in the bottom right figure, where the radius shown is a reference for comparison.  
}
\end{figure}

\begin{figure}
\begin{center}
\includegraphics[width=3.8in]{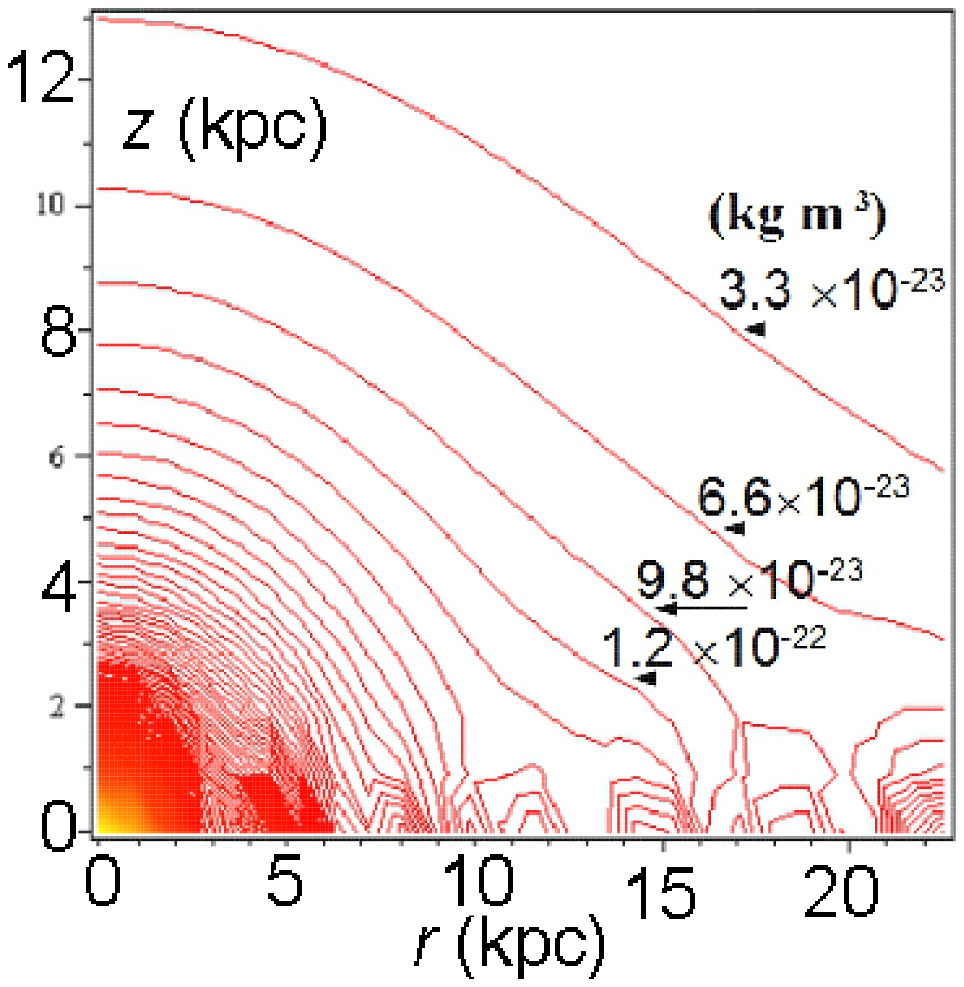}
\end{center}
\caption{
\label{fig:2403.sup}
	  Density contour plot for NGC 2403. The galactic nucleus is positioned at $(r=0, z=0)$.
}
\end{figure}

We studied the galaxy with the aid of several plots and images. Our fit for the RC is shown in  Fig. \ref{fig:RCs}. In Fig. \ref{fig:results2403}  we display the galaxy's density at $z=0$ using a contour plot. Density curves at other $z$-values are illustrated in Fig. \ref{fig:2403.sup}  and they show that the density gradually decays with $z$, as is generally observed in spiral galaxies, a pattern repeated for all the galaxies studied here. In order to have a qualitative means to relate observation to theory, we superimposed a rotated version of  the density contour plot onto a picture of the galaxy\footnote{Picture of NGC 2403 by Fred Calvert$/$Adam Block$/$NOAO$/$AURA$/$NSF, available at $<$http:$//$www.noao.edu$/$outreach$/$aop$/$observers$/$n2403.html$>$.}, as shown in Fig. \ref{fig:results2403}. For the same reason, an infrared picture\footnote{Image available at $<{\rm http}://{\rm ned.ipac.caltech.edu}/{\rm img4}/{\rm 1999A+A...345...59L}/{\rm NGC}\_{\rm 2403}$\\${\rm :I:V:l1999.gif}>$.} of NGC 2403 was combined with the density plot as shown in Fig. \ref{fig:results2403}.

We also calculated the galaxy's mass and determined its optical radius, presented in Table \ref{tab:1}, where observational data are included for comparison. The theoretical optical radius can be visualized in  the logarithmic plot of the density in  Fig. \ref{fig:results2403}   and, based on the hypothesis presented before, it is defined to be the radius at which the density curve first intercepts the $log_{10}(\rho)= -21.75$ threshold. 

	Regarding the RC fit, the plot in Fig. \ref{fig:RCs}  presents an excellent adherence to the data. From that fit alone one could be inclined to conclude that NGC 2403 can be modeled as a low-density dust cloud rotating in stationary, axially-symmetric motion. 
	
As a check, we calculated a RC by applying our GR density to Newtonian theory, using equation A.17 from Casertano\cite{cas}. In the numerical integration we chose to set the integration limit in r up to the last RC data point as after this point, the mass density curve loses accuracy, based as it is on coefficients calculated for a fit up to that last data point. In order to effect improvement, high quality data points at greater radii are required. One consequence of this choice is that for radii close to the last data point, the Newtonian curve may be inaccurate. The  general profile of the curve for radii interior to this point, on the other hand, does not suffer from this limitation.

\begin{table*}
 \begin{minipage}{140mm}
  \caption{
  \label{tab:1}
  Galaxies' masses and optical radii. 
  }
	\begin{tabular}{@{}ccccccc@{}}
  \hline
              &   \multicolumn{3}{c}{Mass ($10^{10}$ M$_{\odot}$)} & \multicolumn{3}{c}{Optical radius (kpc) }\\
 Galaxy       & This & Other &      & This & Other  &  \\
							& work & sources & Ref.\footnote{The columns ``Ref." show the references for the results listed under ``Other sources".} & work & sources & Ref.  \\
 \hline
 NGC 2403 & 5.9 & 1.7 -- 10.2 & \cite{bo,fr,ke} & 9.32 & 9.26 & \footnote{\cite{pat}: this reference was used to collect the data applied to the calculations, as explained in the text.}  \\
 NGC 2903 & 17.2 & 5.5 -- 19.4 & \cite{ke,sm} & 17.36 & 17.17 & \cite{ke,sm} \\
 NGC 5055 & 27.8 & 32.7 & \cite{ke} & 15.31 & 17.65 & \cite{ke,sm}  \\
 Milky Way & 9.3 & $> 12.5$ & \cite{sho} & $> 15$ & 19-21 & \cite{ct07}  \\
\hline
\end{tabular}
\end{minipage}
\end{table*}	

The velocities in the Newtonian RC,  shown in Fig. \ref{fig:RCs}, are far lower than the data points, indicating that our GR mass density function is not sufficient to generate an appropriate RC when Newtonian gravity theory is applied because the Newtonian RC it generates does not reach up to the full range of the higher-velocity level of the data. 

	That our model is a lowest-order approximation to the reality of the physical source is manifested in the ring structure of the density contour plot in Fig. \ref{fig:results2403}, which shows the approximate mass density that generates the rotation curve, i.e, the mass density distribution of the galaxy itself. In that figure the complex structures of NGC 2403 seen in the actual physical image of Fig. \ref{fig:results2403} are idealized into areas of higher and lower densities at certain radii. From this figure, we find a few particular features that are common to the drawing and to the picture:
	\begin{enumerate}
  \item Very bright areas around the nucleus correspond closely to high density areas in the density contour plot in Fig. \ref{fig:results2403}; 
  \item Theoretical outer rings appear located at radii where denser areas of actual arms exist; and
  \item In the drawing, very low density rings are present at radii where the galaxy displays very low brightness. 
\end{enumerate}

	A particular detail is shown in Fig. \ref{fig:results2403}, where the infrared-bright region near the core coincides with the high density region in the theoretical density plot: the densest area extends from the center up to a radius of $2.5$ kpc, followed by a less dense region up to $\sim6$ kpc; beyond this, low-density theoretical rings become the norm. 
	
	As noted earlier, the density in the $z$-direction diminishes steeply as is to be expected for a spiral galaxy. Therefore, not only is the RC very well-fitted by our model, but also the theoretical galactic density provides a reasonable insight into the actual structure of NGC 2403, particularly at radii internal to its minor axis, where the galaxy becomes more circular. 
	
\begin{figure}
\begin{center}
\includegraphics[width=4.2in]{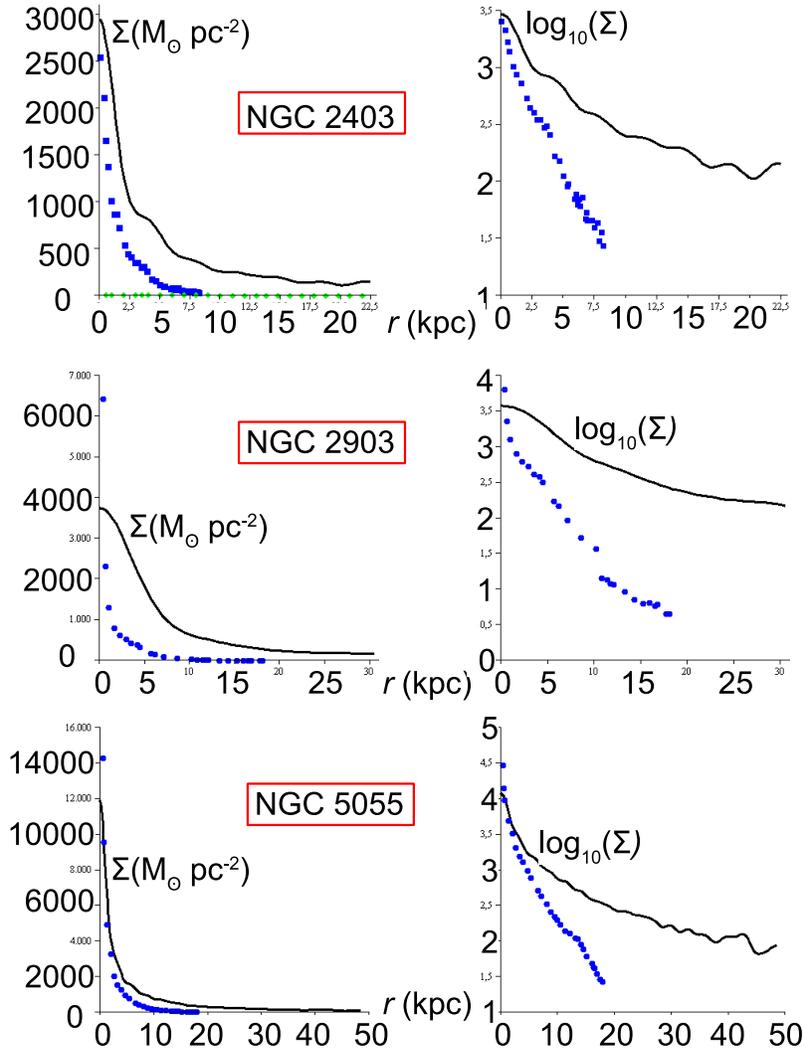}
\end{center}
\caption{
\label{fig:lum}
	The solid lines correspond to surface density curves, $\Sigma$ (in M$_{\odot}$ pc$^{-2}$), on the left, and their respective log10 curves on the right, for the galaxies NGC 2403, NGC 2903 and NGC 5055, calculated with the GR galactic model. The data points in blue correspond to surface brightness in the {\it 3.6} $\mu${\it m}-band (on the left and in L$_{\odot}$ pc$^{-2}$) with the respective log10 points in the figures on the right and they were extracted from plots in \cite{db}.  The scale on the $y$-axes can be used both for the lines and for the data points.
}
\end{figure}

	It is noteworthy that, at the galactic midplane, the density curve shows in Fig. \ref{fig:results2403}, a steep descent near the nucleus. This strong fall-off is an interesting property that arises naturally from the RC data applied to the model, and it repeats in the other galaxies of this study. Its quality resides in the fact that luminosity densities of galaxies also present a strong fall-off and we adhere to the generally-accepted view that galactic luminosity and density are intimately related.  	We investigated this relationship within our model by comparing the galactic surface density as calculated from our density expression with luminosity data from de Blok et al. \cite{db} The result is shown in Figure \ref{fig:lum}.  In order to obtain this figure, we integrated the GR mass density in $z$ up to a value beyond which the surface density did not change significantly. 	The figure on the left shows that the surface density curve follows a contour similar to the surface brightness data, reinforcing the qualitative mapping that was presented above. As well, it shows that $M/L$ is greater than 1, particularly for large radii indicating the existence of matter in the galaxy that is not observed through luminosity data. This is the matter that we previously referred to as excess NLM. In what follows we calculate the total galactic mass as well as the value of the NLM contribution to it, showing that the latter is significantly less than the dark matter mass obtained using NG.

In order to obtain $ M/L \sim 1$ (to help determine the limiting height, z$_{lum}$, beyond which the NLM contribution to the galactic mass becomes relevant), we integrated the mass density in $z$ up to a cut-off height of z$_{lum}$= 978 pc. We obtained this value by numerically calculating the surface density for different z-values until a good visual match was found between the surface density curve and the surface brightness data points. The cut-off z$_{lum}$ is 4.3$\%$ of the radius of the last RC data point, $R = 22.5$ kpc ($z/R = 0.043$) and it signals a limit in the galaxy, in the $z$ direction, where visible light would be observed. Beyond this point matter is expected to contribute little to the observed luminosity.

With regard to the total galactic mass, integration of the mass density function up to the radius of the last RC data point yielded $5.9 \times 10^{10}$ M$_\odot$, a value that, while larger than that derived by MOND \cite{bo}, is considerably lower than others that apply NG \cite{ke,fr}. 

We estimated the mass associated with the luminous part of the galaxy by integrating the mass density function within a disk of radius equal to the last RC data point and with height given by two times z$_{lum}$, finding $1.3 \times 10^{10}$ M$_\odot$, a value close to the total baryonic mass found in the literature \cite{ke,vb}. This implies that the mass of the NLM portion of the galaxy in this GR model is $4.6 \times 10^{10}$ M$_\odot$, a value 37.0\% smaller than the DM portion of the galaxy calculated by Kent\cite{ke} ($7.3 \times 10^{10}$ M$_\odot$).

 	Our result is consistent with the general-relativistic context in the sense that the flattened RC clearly bears the imprint of general relativistic effects that cannot be neglected in the system under consideration, effects that are not present when applying Newtonian gravity, as we discussed in our earlier papers and which we repeat in appendix \ref{ap:1}.

	Our surface density result suggests that there may exist more matter in this galaxy than has been observed thus-far. It is to be noted that the amount of undetected matter indicated by GR is smaller than the  DM which is required to fit the data when Newtonian gravity is the chosen theory of gravity.  This difference merits further investigation.  	Clearly the perfect fluid source in our GR model does not differentiate between stars, dust or gas.  There are ongoing efforts to pursue such identification. For example, regarding the Milky Way, analysis by  Gupta et al.\cite{gu} presents the possibility of the existence of a very large, highly massive reservoir of gas in the Galaxy, extending to over $100$ kpc in galactocentric radius. On the other hand, it is noteworthy that this GR model reduces significantly the amount of NLM necessary to fit the galactic RC if one assumes that both the model and the data are sufficiently accurate.

	Our final note on NGC 2403 refers to the  minimal density hypothesis. Table \ref{tab:1} presents our theoretical value for the optical radius, $r_{opt} = 9.32$ kpc, which corresponds to the radius where the $log_{10}(\rho)$ plot of Fig. \ref{fig:results2403} first has the threshold value $-21.75$. This theoretical optical radius is remarkably close to the one calculated from observational data, namely $9.26$ kpc, also shown in that table. This agreement, further confirming our earlier analyses\cite{ct07,cc12}, reinforces the hypothesis that in this model the density $\rho_{opt} = 10^{-21.75}$ kg m$^{-3}$  is deeply connected to the isophotal limiting surface brightness of $25$ mag arcsec$^{-2}$. Also, Fig. \ref{fig:results2403} shows that the $log_{10}(\rho) = -21.75$ line is crossed several times by the density curve. This suggests that 
the optical luminosity zone for this galaxy could extend to 15 kpc.

\section{NGC 2903}\label{sec:2903}

The spiral galaxy NGC 2903 was included in this sample because it has been studied previously in the literature\cite{cc12,db,ke,sm} and has proven to be challenging to be fitted with MOND\cite{an}.   Also, it has tightly-wound spiral arms as well as a bar and we wished to investigate whether the presence of these features would influence our results.

\begin{figure}
\begin{center}
\includegraphics[width=2.7in]{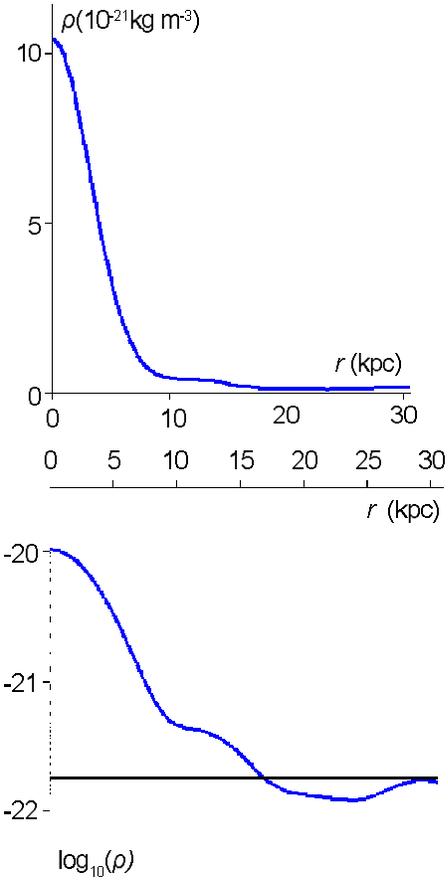}
\end{center}
\caption{
\label{fig:dens2903}
	Volume mass density plots for NGC 2903 at $z=0$. Top: density curve  derived from the expansion of the velocity function, with the density ranging from $\sim 1.2 \times 10^{-22}$ to $\sim 1 \times 10^{-20}$ kg m$^{-3}$. Bottom: logarithmic plot for the density, with the horizontal line at $-21.75$.  
  }
\end{figure}

\begin{figure}
\begin{center}
\includegraphics[width=4.5in]{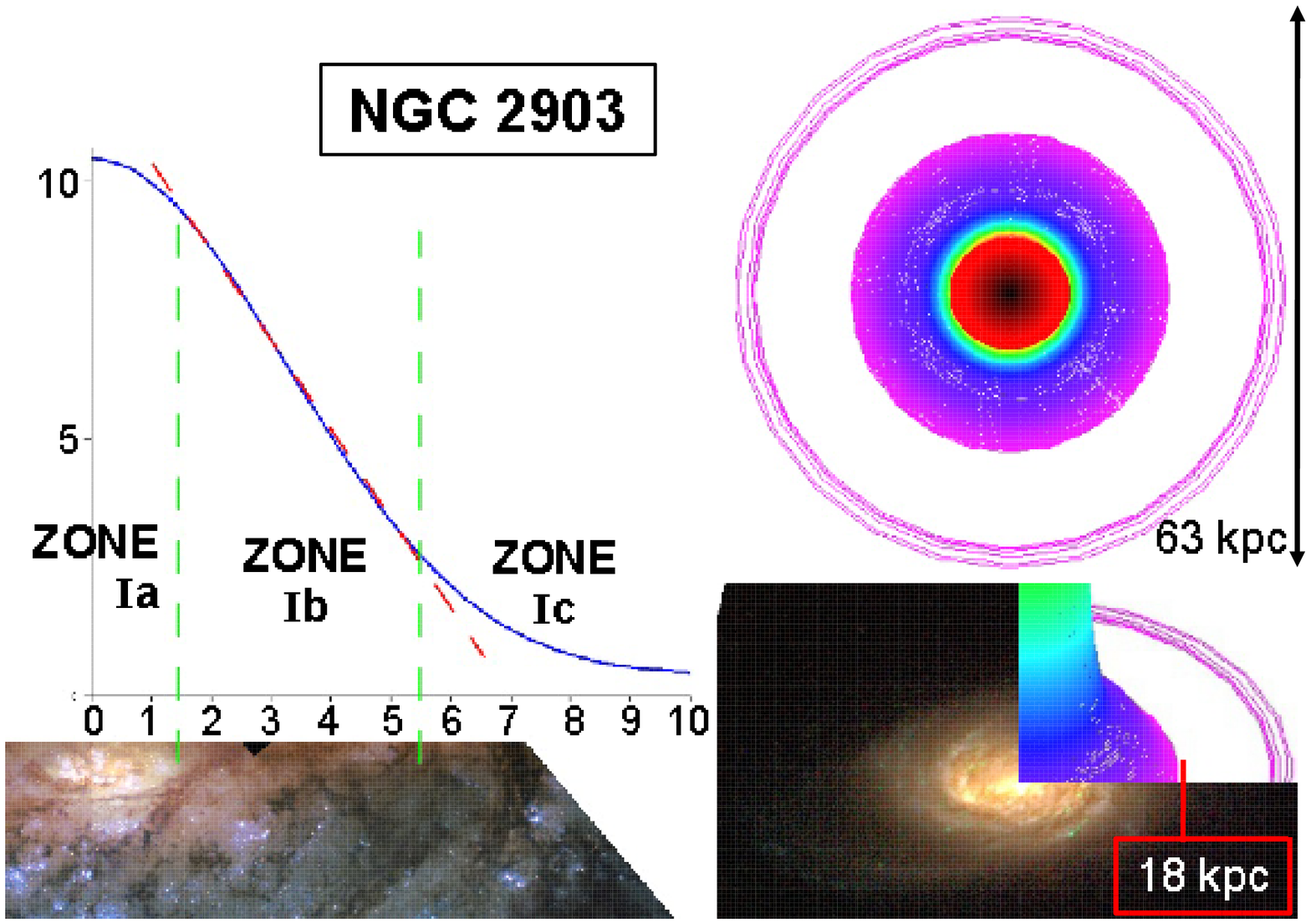}
\end{center}
\caption{
\label{fig:contour2903}
 Results for NGC 2903. On the left, an image of the galactic central area with a section of the density plot (detailed in Fig. \ref{fig:dens2903}) superimposed on it for comparison.  The three zones are discussed in the text. On the top right is the contour plot for the density distribution in the galactic midplane, where the colors follow the sequence of the legend in Fig. \ref{fig:results2403} with the numerical values of Fig. \ref{fig:dens2903}. The bottom right figure presents a combination of the contour plot (rotated) with a picture of the galaxy, where the distance shown is related to the galactic optical radius. 
}
\end{figure}

We present our rotation curve in Fig. \ref{fig:RCs} with $n = 7$ parameters using data from \cite{db}. The mass density plots in Figs. \ref{fig:dens2903} and \ref{fig:contour2903} correspond well to the galaxy's general features as seen in its pictures, such as the presence of a high density area close to the nucleus, evident in Fig. \ref{fig:contour2903}. It is noteworthy that this galaxy is the least dense in our sample.

	We identified two basic zones in the galactic disk. In Fig. \ref{fig:contour2903}, zone I is the high density disk that encompasses the galactic nucleus\footnote{Picture available at $<$http:$/$$/$ned.ipac.caltech.edu$/$img1$/$ 1996AJ....111..174F$/$NGC$\_$2903:I:gir:fgg1996.jpg$>.$}. At a radius near 18 kpc, zone II begins and it extends up to the galactic edge, at which point the density is remarkably low. 
	
	Since there is available a zoomed image of this galaxy near the nucleus\footnote{Image available at $<$http:$/$$/$apod.nasa.gov$/$apod$/$image$/$0103$/$ ngc2903$\_$hst$\_$big.jpg$>$.}, we zoomed the density plot of Fig.  \ref{fig:dens2903} for the first half of zone I, as shown in Fig.  \ref{fig:contour2903}. We noted that for radii between 1.5 kpc and 5.5 kpc (zone Ib) the density decays linearly, perhaps suggesting a region with a particular dynamical behaviour. As in the case of NGC 2403, the Newtonian curve in Fig. \ref{fig:RCs}  never reaches up to the data.

The theoretical density zones appear to match the images of the galaxies. Due to the rotational symmetry of the model however, the bar structure of this galaxy could not be gleaned from this model.

	We compared the galactic surface density calculated as explained in the previous section, from our density expression with luminosity data from reference \cite{db}, the result shown in Fig. \ref{fig:lum}. All considerations given thus-far regarding the surface density of NGC 2403 apply to this galaxy's surface density as well. We integrated the mass density in $z$ up to a cut-off height z$_{lum}= 437$ pc to obtain $ M/L \sim 1$ for radii between 1.5 and 10 kpc.  The cut-off z$_{lum}$ is  1.4\%  of the radius of the last RC data point, $R = 30.6$ kpc ($z/R = 0.014$). This ratio is smaller than the one found for NGC 2403, perhaps due to the fact that NGC 2903 is on average less dense than that galaxy. The total mass of this galaxy, integrating the mass density function up to the radius of the last RC data point, yielded $17.2 \times 10^{10}$ M$_\odot$, a value larger than one derived using MOND \cite{sm}, but lower than that which applies Newtonian gravity\cite{ke}. Regarding the latter, in that paper the total mass was obtained integrating up to a radius of 24 kpc; for the sake of comparison, the integration of our mass density function up to $r= 24$ kpc yields a total mass of $14.2 \times 10^{10}$ M$_\odot$. 

	The mass associated with the luminous part of the galaxy, calculated as described in the previous section up to the radius of the last RC data point, was found to be $1.5 \times 10^{10}$ M$_\odot$ (when calculated up to $r = 24$ kpc the luminous mass was $1.3 \times 10^{10}$ M$_\odot$) . The value is compatible with the one found by \cite{ke}, considering that this author assumed $ M/L \sim 3.35$ while we focused on a close match between surface brightness and surface density (i.e., $ M/L \sim 1$) in a range of radii closer to the galactic center. Our result for the luminous mass implies that the mass of the NLM portion of this galaxy, assuming a galactic radius of 24 kpc, is $12.9 \times 10^{10}$ M$_\odot$, a value 20.4\%  smaller than the DM portion of the galaxy calculated by Kent\cite{ke} ($16.2 \times 10^{10}$ M$_\odot$). Had we  considered a mass-to-light ratio higher than 1 as did Kent, the luminous mass would have been larger than what we found and, consequently, the galactic NLM  would have had a mass even lower than what which we presented above, increasing the latter percentage.

	The galactic mass and radius which we found for NGC 2903 are shown in Table \ref{tab:1}. They reinforce our conclusions drawn from the results for NGC 2403. As well, the density $\rho_{opt}= 10^{-21.75}$ kg m$^{-3}$ continues to reveal its connection to the isophotal limiting surface brightness of 25 mag arcsec$^{-2}$. In this case the logarithmic plot in Fig. \ref{fig:dens2903} shows that the landmark $r_{opt} = 17.36$ kpc clearly separates a high density region from a low density one. The image of NGC 2903 in Fig. \ref{fig:contour2903} shows that these regions are indeed evident, a situation different from the one posed by NGC 2403.

\section{NGC 5055}\label{sec:5055}

The Sunflower galaxy, NGC 5055, had its rotation curve studied previously \cite{ke} and has more recent observational data available \cite{db}, which we used to obtain our rotation curve, shown in Fig. \ref{fig:RCs}. We chose $n = 37$ parameters for the expansion in equation (\ref{eq2}) for a good RC fit, particularly in the region around 5 kpc, where the observational data show a dip. A larger number of parameters is invariably required to capture particular features near the origin, which in turn yield more detailed information regarding the mass density distribution.

\begin{figure}
\begin{center}
\includegraphics[width=2.8in]{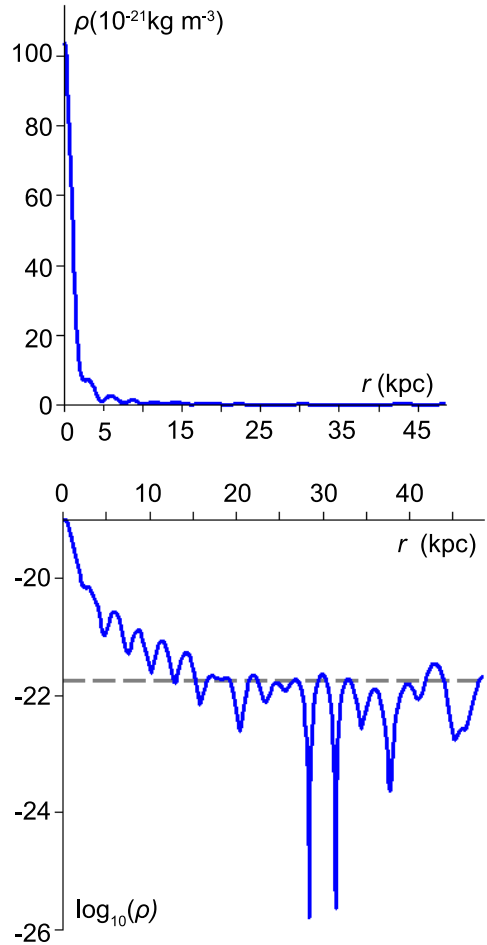}
\end{center}
\caption{
\label{fig:dens5055}
Volume mass density plots for NGC 5055 at $z=0$. Top: density curve  derived from the expansion of the velocity function, with the density ranging from $\sim 1.4 \times 10^{-26}$ to $\sim 1 \times 10^{-19}$ kg m$^{-3}$. Bottom: logarithmic plot for the density, with the dashed line at $-21.75$.  
 }
\end{figure}

\begin{figure}
\begin{center}
\includegraphics[width=3.9in]{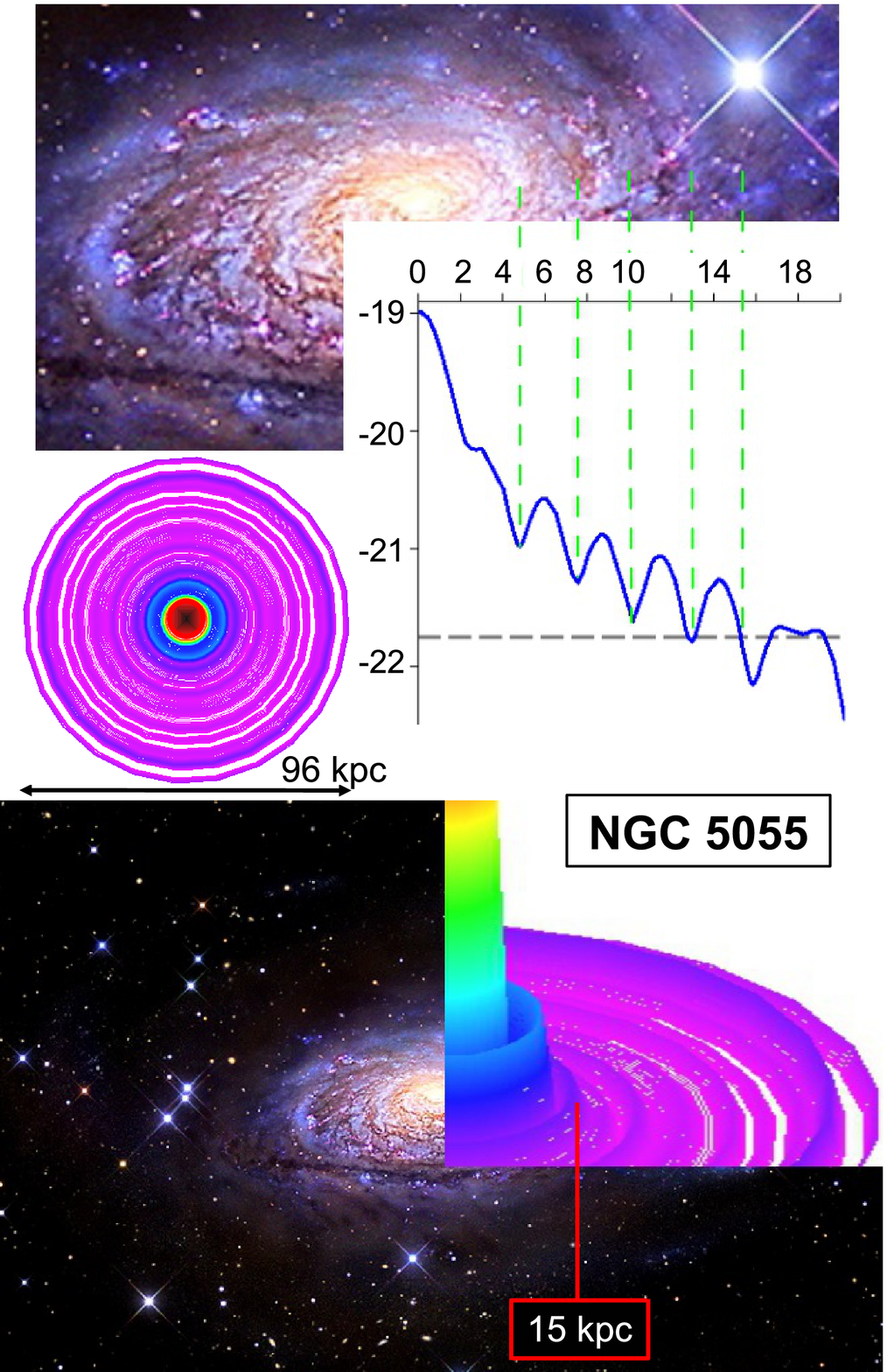}
\end{center}
\caption{
\label{fig:contour5055}
	Results for NGC 5055. Top: an image of the galactic central area with a section of the density plot (detailed in Fig. \ref{fig:dens5055}) superimposed to it for comparison.  Middle: contour plot for the density distribution in the galactic midplane, where the colors follow the sequence of the legend in Fig. \ref{fig:results2403}  with the numerical values of Fig. \ref{fig:dens5055}. Bottom: a combination of the contour plot (rotated) with a picture of the galaxy, where the distance shown is related to the galactic optical radius. 
}
\end{figure}

The density plots in Figs. \ref{fig:dens5055} and \ref{fig:contour5055} match the galaxy's general features as can be seen in comparison to its images. 	In the picture\footnote{The picture by R. Jay GaBany is available at $<$http://apod.nasa.gov/apod/ap100911.html$>$.} of the bottom image of Fig.  \ref{fig:contour5055} the high density area within the radius of 15 kpc in the vicinity of the nucleus is matched by our theoretical density plot displayed above  it, as well as the low density region that goes beyond it. 

	The central core of this galaxy presents several fragmented arms as displayed by its pictures. These density variation areas can be theoretically associated, on average, with the density variations in the contour plot of Fig. \ref{fig:contour5055}. In the bottom part of this figure we relate our result to the galactic image using the theoretical maxima as references. The correspondence between rings and arms is necessarily an approximation since the rings are the result of averaged densities along the radii due to the imposed axial symmetry of the model. The overall variation of the density function with the radius however, is a relevant piece of information. For instance, the steep decrease in the density function from the nucleus until $r\sim 5$ kpc (coincidentally, where the dip in the RC is located) suggests a very high density region with physical properties that are different from the rest of the galaxy. 
	
	The Newtonian RC in Fig. \ref{fig:RCs}, that we obtained using the GR density, did not reach up to the RC data points, as occurred with the preceding galaxies, demonstrating the incompatibility of the GR mass density with NG.  

 We compared the galactic surface density, calculated as explained before, from our density expression with luminosity data from \cite{db}, a result also shown in Fig. \ref{fig:lum}. We integrated the mass density in $z$ up to a cut-off height z$_{lum}= 2430$ pc to obtain M/L$\sim 1$.  The cut-off z$_{lum}$ is  5.0\%  of the radius of the last RC data point, $R = 48.5$ kpc ($z/R = 0.05$). This ratio is higher than the one found for NGC 2403, perhaps due to the fact that NGC 5055 is on average, denser than that galaxy.

The total mass of this galaxy, obtained by integrating the mass density function up to the radius of the last RC data point, yielded $27.8 \times 10^{10}$ M$_\odot$. This value is smaller than one that applies Newtonian gravity using data up to a smaller radius of 45 kpc \cite{ke}, which is $32.7 \times 10^{10}$ M$_\odot$. The total mass of the galaxy that we obtain by integrating our mass density function up to $r = 45 $ kpc is $ 26.9 \times 10^{10}$ M$_\odot$.

The mass associated with the luminous part of the galaxy, calculated assuming M/L $\sim$ 1 as described in the NGC 2403 Section (integrated up to the radius of the last RC data point), was found to be $\sim 8.5 \times 10^{10}$ M$_\odot$.  

 	For the sake of comparison we calculated the mass associated with the luminous part of the galaxy up to a radius of 45 kpc, finding $ 8.5 \times 10^{10}$ M$_\odot$. The corresponding value found by Kent\cite{ke}, assuming  M/L $\sim$ 3, was $8.2 \times 10^{10}$ M$_\odot$. 	Our result for the luminous mass implies that the mass of the NLM portion of this galaxy up to a galactocentric radius of 45 kpc is $18.4 \times 10^{10}$ M$_\odot$, which is $\sim$25\% smaller than the DM portion of the galaxy calculated by Kent ($24.5 \times 10^{10}$ M$_\odot$). Had we assumed a higher mass-to-light ratio as did Kent, we would have obtained an even smaller mass for the NLM portion of this galaxy. 

From the minimum density threshold we determined the galactic optical radius as r$_{opt} =$ 15.31 kpc. This landmark is  87\% of the value of the optical radius calculated from observational data, presented in Table \ref{tab:1}.  In fact, from the logarithmic plot in Fig \ref{fig:dens5055} we see that at $r = $ 17.65 kpc (the calculated optical radius) we also have log$_{10}\rho \sim -21.75$, suggesting a gradually fading optical zone.

\section{The Milky Way}\label{sec:mw}

Our Galaxy's RC has the feature that near the nucleus, stars apparently have non-zero velocities, as the observational data\cite{sho} show (the red circles in the RC of Fig. \ref{fig:RCs}). This was a new challenge to our model because the Bessel functions that we use in the expansion necessarily start with $v = 0$ km s$^{-1}$ at $r = 0$ kpc. Nevertheless, our model presented remarkable results, as we show below.

	Like the RC of NGC 5055, also a barred galaxy, the Milky Way's RC presents a steep rise close to the nucleus, reaching a maximum near a radius of $0.3$ kpc. This suggested the requirement for a few tens of expansion coefficients as was the case with NGC 5055. Our Galaxy's RC declines after $r \sim 0.3$ kpc until a radius near $2.5$ kpc, then it grows gradually until a maximum near 6 kpc, from which it starts to fall off, having a minimum at $9 \to 10$ kpc; then the curve grows again until nearly $13$ kpc. Beyond this radius, error bars become increasingly larger but it was argued in the literature that the curve reaches a maximum near $15$ kpc and then it declines \cite{so}. All of these features are present in our theoretical curve accompanied by  some wavy behavior in the curve, as would have been predicted from the result for NGC 5055 in Fig. \ref{fig:RCs}. In general, the more oscillatory the velocity data, the more terms that are needed in the Bessel expansion for the theory to reproduce those oscillations. The oscillatory level naturally manifests itself as rings on the mass density distribution that the theory provides. This relation implies that the rings present in that distribution have physical origins, which are related to the number of terms, $n$, of the expansion.

	An important observational feature that had to be met by our model, as pointed out in \cite{fp}, is the value of the local mass density in the vicinity of the Sun of $\rho(r_{\odot} , z=0) = 6.3 \times 10^{-21}$ kg m$^{-3}$ . We used this value as one of the constraints to select the number of parameters in the expansion, the other being to fit the RC that displayed the basic features described above. 

We obtained the rotation curve in Fig. \ref{fig:RCs} using $n=29$ parameters. The wavy nature of the curve is present as expected and its general behaviour follows closely the features described above. In particular, the peak at $r \sim 0.3$ kpc is present. Once more, the Newtonian RC displayed in that figure does not reach up to the data. 

\begin{figure}
\begin{center}
\includegraphics[width=2.8in]{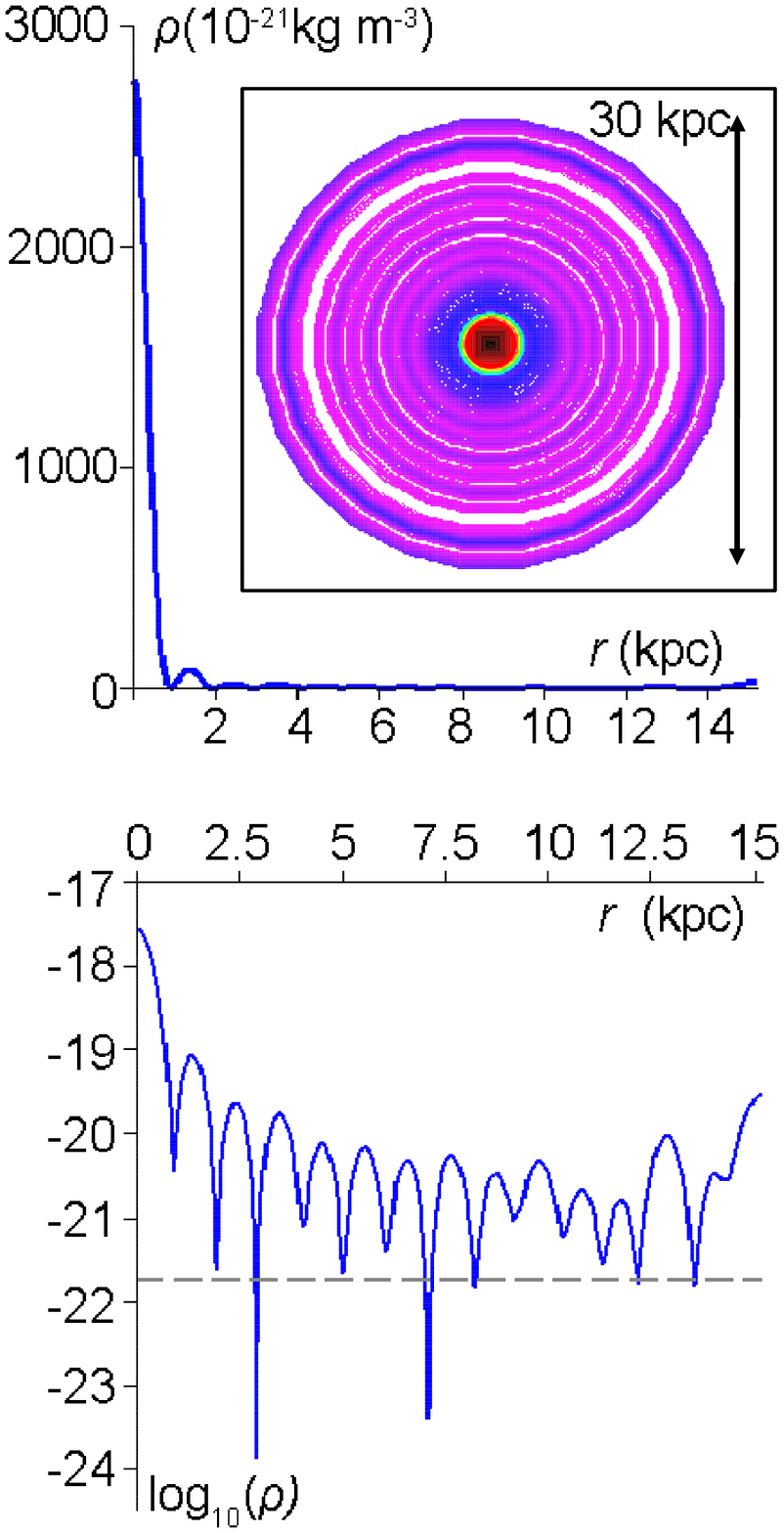}
\end{center}
\caption{
\label{fig:densmw}
 Volume mass density plots for the Milky Way at $z=0$. Top: density curve  derived from the expansion of the velocity function, with the density ranging from $\sim 1.3 \times 10^{-24}$ to $\sim 2.6 \times 10^{-18}$ kg m$^{-3}$. In the box  the contour plot for the density distribution is displayed, with the colors following the sequence of the legend in Fig. \ref{fig:results2403}  with the numerical values of the density curve shown here.
Bottom: logarithmic plot for the density, with the dashed line at $-21.75$.  
}
\end{figure}

	The mass density plots are shown in Fig. \ref{fig:densmw}, displaying a very high density core, the densest among the galaxies that we studied, and concentrated within only 1 kpc from the Galactic nucleus, after which the first ring is evident. As for the local mass density near the Sun, we obtained a value that was close to the observational one, reaching $5.8 \times 10^{-21}$ kg m$^{-3}$ at 7.66 kpc. This radius is well within the range of the Sun's position in the Milky Way, whose distance to the Galactic center is estimated \cite{ch} to be between 7.3 and 8.5 kpc.

	An evident characteristic of the logarithmic plot of Fig. \ref{fig:densmw} is that the mass density does \textit{not} decay significantly to values less than $10^{-21.75}$ kg m$^{-3}$. From the analysis performed in the same kind of plot for other galaxies, we are led to interpret this result, based on the data set used, as the optical radius of our galaxy being \textit{beyond} 15 kpc. Therefore a significant number of stars is required to be observed at large radii, and as accurately as possible, for the Galactic optical radius to be better-determined with our method. Our limit is consistent with our previous estimate of an optical radius of $\sim 19-21$ kpc, listed in Table \ref{tab:1}, which used a smaller number of velocity points.
	
The average density distribution of mass in the Galactic midplane is shown in the contour plot of Fig. \ref{fig:densmw}, where the Galactic disk has a diameter of 30 kpc, which is the commonly accepted value for its actual diameter\cite{rcm}, prior to the new results of Xu et al.\cite{xu} This figure has patterns similar to those of the contour plot of Fig. \ref{fig:contour5055}, suggesting that, based on the data available thus-far, the Milky Way would appear to a distant observer roughly as NGC 5055 appears to us. To our knowledge, such a theoretical map of the Galaxy was never previously presented in the literature. We submit that it will contribute to a better knowledge of the structure of our Galaxy, a knowledge base which is still rather limited\cite{ch}. 

\begin{figure}
\begin{center}
\includegraphics[width=3in]{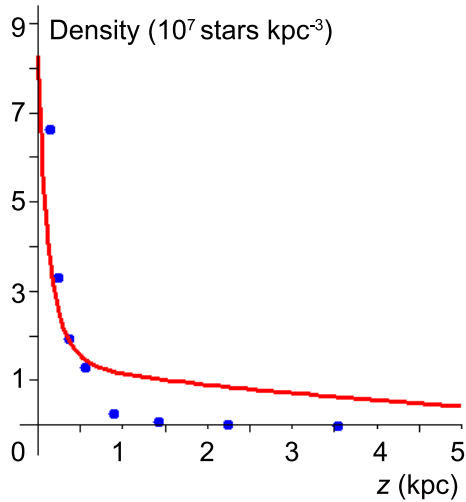}
\end{center}
\caption{
\label{fig:mw.5}
			The solid curve shows our result for the number of stars at a possible position for our Sun ($r =7.66$ kpc) as a function of $z$, the distance from the Galactic midplane. The dots show observational data points at the Sun's vicinity. 
}
\end{figure}

	Figure \ref{fig:mw.5} shows the dependency of the density with the distance z to the Galactic midplane for $r = 7.66$ kpc. Our result is the solid line while the dots show actual data \cite{fp}. Our curve fits well the data up to $z \sim$ 0.5 kpc. Higher distances are more challenging to fit in this case because, as mentioned above, the coefficients of the expansion were calculated for $z=0$ and also because of contributions from NLM at higher $z$.

	We calculated the mass of the Galaxy as $9.3 \times 10^{10}$ M$_{\odot}$ within the data range of $15$ kpc. This value is close to the one estimated for the contributions due to bulge, disk and arms ($8.3 \times 10^{10}$ M$_{\odot}$) in reference \cite{sho} but is still $\sim$ 10\% larger, this extra mass being of unknown origin. One might argue that within the context of astrophysical accuracy, our value is consistent with the contributions from bulge, disk and arms alone.
	
Sofue et al.\cite{sho} had to assume an additional mass contribution from DM in order to fit the Galactic rotation curve, which significantly increased their value for the total mass of the Milky Way (total mass of 12.5 $\times$ 10$^{10}$ M$_{\odot}$ within a radius of 10 kpc and of 20.4 $\times$ 10$^{10}$ M$_{\odot}$ within a radius of 20 kpc), shown in Table \ref{tab:1}. \textit{The need for DM in this case is between 3 and 6 times the need for NLM in the GR model.}

In \cite{ct07}, working with a smaller number of the then available data points for the Milky Way but using data with radii as far as the 30 kpc range, we witnessed the following for the log 10 density plot (in Figure 7.a of \cite{ct07}) :
\begin{enumerate}
\item a first crossing of the threshold log 10 density value of -21.75 at a radius of 19 kpc followed by
\item a very close hovering below -21.75 with a re-crossing of the threshold value at 21 kpc followed by 
\item an almost equally close hovering above the threshold value, re-crossing the threshold value at a radius of approximately 23 kpc.
\item Beyond 23 kpc, the dip below the threshold density value was substantial.
\end{enumerate}
Thus, while we reported then an optical radius range of 19-21 kpc, arguably it could have been registered as 19-23 kpc, based on the above.
In the present work, we had the advantage of access to considerably more data and we concentrated on fitting our curves to match this data. We focused on the data up to a radius of 15 kpc as the now larger store of available data beyond 15 kpc was rather erratic and not sufficiently meaningful to attempt a fit. However, using the reliable range up to 15 kpc, we note from Figure \ref{fig:densmw} that apart from two very narrow (and hence insignificant) spikes below the threshold density, the log 10 density plot for the Milky Way remains above the threshold density value through the entire range up to 15 kpc. Based upon the experience from the other galaxies, it is thus clear that the optical radius of the Milky Way lies considerably above 15 kpc. Hopefully a greater number of reliable data will become available in the future to extend the detailed rotation curve beyond 15 kpc to enable us to ascertain the value of the threshold crossing point. 

	Assuming that the \textit{optical} radius of the Galaxy lies in the neighbourhood of 20 kpc, it is possible, based on the general pattern of the other galaxies, that its \textit{total} radius lies near $30$ kpc. We predicted this larger radius for the Galaxy in communications to conferences in 2013 (20$^{th}$ GRG) and 2014 (NEB 16). The very recent publication by Xu et al.\cite{xu} includes observational evidence that our Galaxy has a radius of at least 20 kpc. This confirmation of our theory-based prediction lends confidence to the present GR model in the description of galactic dynamics.
	
	Assuming that the Milky Way has a 30 kpc radius, our 29-parameter fit yields a total mass of $22.2 \times 10^{10}$ M$_{\odot}$ while using dark matter in the context of the model by Sofue et al.\cite{sho},  we found a total galactic mass of nearly $33 \times 10^{10}$ M$_{\odot}$ for this radius. This indicates that in this case the amount of NLM in the GR context is 33\% less than the amount of DM under NG.

\section{Concluding remarks}\label{sec:con}

The rotation curves for the galaxies generated with our GR model display good fits to the data, which would have been impossible if the Newtonian approach were used with our general relativistic mass densities, as we showed. Our model is based upon the assumption of a continuum of pressure-free fluid in perfectly axially symmetric distribution about an axis of rotation. The reality of a galaxy is that it is only, to a highly varying degree depending upon the galaxy chosen, approximately axially symmetric, although this approximation has yielded many good results in studies of galaxies. Moreover, it is not a continuum but rather consists of concentrations of mass in the form of stars.  These two factors alone must necessarily alter the gravitational field being idealized and hence impose \textit{a priori} limitations on what can be expected from our model. In spite of this, the concordance with reality that we have witnessed is rather remarkable, considering its inherent limitations. 

The GR model can be improved in several ways: first, using a metric that includes more detailed structure of the galaxy, such as bulge and arms; in the case of the bulge it would be interesting to consider other star velocity patterns, such as the cylindrical rotation pattern which apparently occurs in the Milky Way\cite{kun}. 
Second is the use of a particulate structure for the galaxy rather than a fluid, for which we foresee a massive use of numerical computations.

Third could be the employment of other solutions instead of  (\ref{eq:a10}), as proposed by Balasin and Grumiller\cite{bg}. However, even with the solution that we use, other improvements could be applied such as the use of values of $z$ other than zero in the determination of RCs.  Related to symmetry constraints, while we have $r$ and $z$ dependence in our functions, $z$ dependence in data points is not recorded and $\phi$ dependence is not present in our functions and not present in the data. While the lack in $z$ dependence might be seen as not very important (the disks are fairly compactified), the lack of $\phi$ dependence means that we are performing an averaging which is also weighted by our data sampling, according to the $\phi$  sector that is most readily accessible, particularly for observations from our position in the Milky Way.	

	Of particular note, the theoretical density zones appear to match the images of the galaxies. This leads us to propose that our method presents a practical first-order mapping of galaxies.  As well this could be viewed as a test of GR beyond the solar system.
	
		The model also allows the determination of the galactic optical radius since the working hypothesis introduced in an earlier paper was confirmed with the galaxies investigated here. We suggest that the effectiveness of the model in this aspect may be physically tied to the brightness of molecular clouds in galaxies. 
	
	Moreover, the indications from the modeling point to the Milky Way having a larger diameter than is generally admitted, a result that was derived from a mass density function that provided good concordance with the local mass density determined by \cite{hf}. This concordance is one of the improvements that this work makes to \cite{ct07}.

	Also regarding the Milky Way, as new, more accurate data points become available for radii beyond $8$ kpc, we expect that our rotation curve will present an even a better fit, particularly for larger radii. Our good matching of the Milky Way rotation curve led to a predicted density of matter in our solar neighbourhood that is close to the observed density. On the other hand, Newtonian gravity modeling still requires some DM even at our position in the Galaxy\cite{so}.  
	
By investigating the galactic M/L, we concluded that a significant amount of yet-unseen matter (NLM) is expected to be present in galaxies. We calculated this fraction and compared it to the fraction of DM estimated using Newtonian gravity. We found that the amount of NLM is significantly less than the amount of DM. 

Therefore, while at this stage we cannot know the precise nature of the NLM that we have deduced to be present, it is considerably more modest in extent than the DM extent claimed on the basis of Newtonian gravitational dynamics. This leads us to feel more comfortable with the hypothesis that it will ultimately be shown to be normal matter within the Standard Model of particle physics.

Finally, the indications from the modeling point to the Milky Way having a larger diameter than is generally admitted, a result that was derived from a mass density function that provided good concordance with the local mass density\cite{hf}. This prediction is one of the several improvements that this work makes to \cite{ct07} and is a result now observationally verified by the recent paper by Xu et al.\cite{xu}

\section*{Acknowledgments}

N.S.M. acknowledges support from C. Frajuca, L. V. Rizzo, M. A. F. Dias, J. I. Kim and INCT-A (Brazil) as well as from the Brazilian funding agency CNPq for the grant 241032/2012-1.  The authors thank S. Tieu for providing an initial maple code and technical assistance, R. M. Marinho, Jr. for support with Maple and Maxima, and F. D. A. Hartwick for astronomical information. Most of the computations in this paper were performed using maple(TM), a trademark of Waterloo Maple Inc. We acknowledge the usage of the HyperLeda database (http://leda.univ-lyon1.fr). 


\appendix

\section{Summary of the galactic model}\label{ap:1}

Following \cite{ct07}, in a first-order approximation, a spiral galaxy can be envisioned as a pressure-free fluid composed of a continuum of particles rotating at a time-independent rate around an axis, primarily concentrated on a disk oriented perpendicularly to this axis \cite{at}. In future studies, higher-order approximations could provide further details beyond this picture, including the description of spiral-arm structures where matter accumulates, emanating from regions near the galactic rotation axis. In our present first-order approximation to a galaxy we will also assume that matter is distributed axially-symmetrically. These constraints encompass a degree of clumping of matter in the form of rings which, in more detailed future studies, may evolve into the arms which we actually observe. The absence of pressure is physically justified by the usually very low matter densities present in the galactic environment. 

	From the outset, we apply general relativity, having seen in earlier work \cite{ct07} that the non-linearities come into play in the description of the galactic dynamics for a continuum of particles in free-fall.
	
	Many years ago it was shown in \cite{vs}  that the spacetime in the interior of an axially-symmetric distribution of particles rotating in stationary motion around an axis of symmetry had a particular metric form\footnote{ The metric form we refer to is in equation (3.3) of the paper by \cite{vs}. This type of metric was first studied by Lanczos (Lanczos C., 1924, Zeitschrift f�r Physik, 21, 73). This paper is translated (by C. Hoenselaers) in: Lanczos K., 1997, Gen. Relat. Grav., 29, 363; it is also related to the article: Krasinski A., Gen. Relat. Grav., 1997, 29, 359.}. We adopt this metric here and rewrite it conveniently as  
\begin{equation}
ds^2 = -� e^{\nu} (dr^2 + dz^2) -� (r^2 -� N^2) d\phi ^2 -� 2 N c d\phi dt + c^2 dt^2,
\label{eq:a1}
\end{equation}
where $v$ and $N$ are functions of the cylindrical polar coordinates $r$ , $z$. The cylindrical polar angle in this system is $\phi$ and $c$ is the speed of light in vacuum. We adopt the convention $(+ �- -� -�)$ for the Minkowski metric, with spacetime indices running from 0 to 3 so that $x_0=ct$, $x_1=r$, $x_2=\phi$ and $x_3=z$. 

	Because of the stationary, axially-symmetric constraint, particles will only display different angular velocities for different values of the $r$ and $z$ coordinates. Since we are interested in studying variations in stellar linear velocities with their distances to the galactic center, we first determine those angular velocities.
	
	By adopting a system of reference relative to which the matter is at rest, i. e., to a system that is comoving with the matter, the actual angular velocity, $\omega(r, z)$, of a particle located at a given position $(r, z)$ relative to the local non-corotating frame can be found by locally diagonalizing the metric, yielding
\begin{equation}
\omega = \frac{cN}{r^2 �- N^2}.
\label{eq:a2}
\end{equation}

In the limit $N/r << 1$, which is the case for the low density fluids that we are dealing with, the local angular velocity admits the simple form 

\begin{equation}
\omega (r, z)  \approx  \frac{c N(r, z)}{r^2}. 
\label{eq:a3} 
\end{equation}                                                                                                                
Equation (\ref{eq:a3}) is the link between the key metric function $N(r, z)$ and the actual rotational motion of the fluid. 

It is to be noted that with the large galactic sizes involved, we treat the stars as well as the constituents of the intergalactic gas as particles as a first approximation. A star's tangential velocity, $V$, is given in the non-rotating frame by
\begin{equation}
V(r, z) = r \omega(r, z) \approx \frac{c N(r, z)}{r}.   
\label{eq:a4}
\end{equation}

This expression shows that from the knowledge of $N(r, z)$ we can determine a galaxy's rotation curve (the star's velocity as a function of its galactocentric distance), for different heights, $z$, relative to the galactic midplane. We obtain this function as well as information on the galaxy's mass density with the aid of Einstein's equations.

	For the system under the above conditions, and with terms retained up to the order of the first power of $G$ (Newton's gravitational constant; \cite{ti} presents details on this approximation criterion), Einstein's equations yield the following expression for N:
\begin{equation}
\frac{N_r^2 + N_z^2}{r^2} = \frac{8\pi G \rho}{c^2} , 
\label{eq:a5}     
\end{equation}                                                       
where the subscripts $r$ and $z$ denote partial differentiation with respect to the respective coordinates. This equation reveals the non-linear relation between the pressureless, stationary, freely gravitating rotational motion (embodied in the field $N$) and the mass distribution (represented by $\rho $).

	Einstein's equations also give
\begin{equation}
N_{rr} + N_{zz} �- \frac{N_r}{r} = 0.
\label{eq:a6} 
\end{equation}                                                          
Solving this equation yields the $N$ function required for the determination of the rotation curve of the galaxy from equation (\ref{eq:a4}). 
	
By defining
\begin{equation}
 \Phi = \int{\frac{N}{r}dr},  
 \label{eq:a7}
\end{equation}
equation (\ref{eq:a6}) can be expressed in the form of Laplace's equation
\begin{equation}
\Delta\Phi = 0,     
\label{eq:a8}
\end{equation}
where $\Delta$ is the flat-space Laplacian operator. This shows that $\Phi$, the generator of the $N$ field by definition, is composed of flat-space harmonic functions, which also relate to the stars' velocities, according to equation (\ref{eq:a4}), as 
\begin{equation}
V = c \frac{\partial \Phi}{\partial r} .
\label{eq:a9}
\end{equation}
The determination of the rotation curve is tied to the determination of $\Phi$. 

	Since equation (\ref{eq:a8}) is linear, we can choose a linear superposition of harmonic solutions as
\begin{equation}	
\Phi = \sum_n C_n e^{�- k_n |z|} J_0 (k_n r),
\label{eq:a10}
\end{equation}
where  $J_0(k_r)$ are the zeroth-order Bessel functions and the $C_n$ are constants dependent upon the galaxy's characteristics. In our analysis we use equation (\ref{eq:a10})  in the $+z$ direction alone and, to include the whole range in $z$, we mirror the $+z$ result  into the $-z$ region, following the natural simplifying approximation of galactic reflection symmetry relative to the $z=0$ plane.
 
	Bessel functions of integer order are common in many problems which are solved in cylindrical polar coordinate systems. They are solutions to Bessel's equation, which arises when finding separable solutions to Laplace's equation	\cite{aw}.  The constants $k_n$ are the zeroes of $J_0$ at the range of integration and are calculated from the orthogonality relation that these functions must satisfy. They are readily obtainable in computational packages such as maple.
	
	A solution like equation (\ref{eq:a10}), presented in the literature previously, focused on solving Einstein's equations with Lanczos' metric \cite{dw} but with motivations other than the rotation curve problem. Here the linear superposition given by the summation over a number of parameters, $n$, allows us to calculate $\Phi$ up to the level of accuracy required with rotation curve matching as the focus. 
	
	The solution (\ref{eq:a10}) applied to equation (\ref{eq:a9}) yields
\begin{equation}
V = - c \sum_n k_n C_n e^{�- k_n |z|} J_1 (k_n r), 
\label{eq:a11}
\end{equation}
where the $J_1$ are the first-order Bessel functions. This is the basic equation that we used to obtain our results. By applying real data points $(r, V)$ to equation (\ref{eq:a11}) assuming $z = 0$ (invoking the standard approximation of positioning the stars in the galactic midplane), we found the corresponding $C_n$-coefficients through the use of the \verb"bestfit" function of maple. These coefficients were then substituted back into equation (\ref{eq:a11}) up to a chosen number of parameters and the resulting equation yielded an expression for $N(r, z)$ using equation (\ref{eq:a4}). Finally, the galaxy's density was calculated from equation (\ref{eq:a5}), determining, by volume integration, the galaxy's total mass. In the density plots we noted that there is an increase in density at the largest radius, which is a computational feature at the boundary rather than a physical characteristic of the object.

\section{Calculating the optical radius from observational data}\label{ap:2}

We wrote a code in Maple to calculate the optical radius of a galaxy from catalogued data. We present below an application of the code to NGC 2403 using data from Hyperleda \cite{pat}.

\begin{verbatim}
# Galactic distance from the Sun, 
# in distance modulus (mag):
> mu:= 27.52: # From Hyperleda
# Galactic distance from the Sun, in parsecs:
> dpc:=10^(1+mu/5);
                               3.191537855e10 
# "log25" (decimal logarithm of apparent diameter,
# in log10(0.1*arcmin)):
> log25:=2.3: # From Hyperleda
# Apparent diameter in arcmin:
> d_ap_arcmin:= (10^(log25))*0.1;
                                 19.95262315
# Apparent diameter in arcseconds:
> d_ap_arcsec:= evalf(d_ap_arcmin*60);
                                 1197.157389
# Galactic optical diameter in kpc:
> diam_opt:= (1/206265)*d_ap_arcsec*dpc/1000;
                                 18.52361343
# Galactic optical radius in kpc:
> r_opt:= diam_opt/2;
                                 9.261806715

\end{verbatim}

\section{Data sets used}\label{ap:3}

All the data sets that we used to obtain our results were extracted from the galaxies' corresponding rotation curve plots in the respective references. The data set for NGC 2403 was \cite{db, fr}:

\begin{verbatim}
[NGC 2403, galactocentric distances, in kpc] = 
[0.07,0.12,0.19,0.25,0.31,0.37,0.44,0.50,0.56,
0.63,0.69,0.75,0.81,0.98,1.21,1.29,1.37,1.49,
1.55,1.63,1.68,1.78,1.92,2.02,2.28,2.15, 2.41,
2.64,2.75,2.97,3.09,3.21,3.31,3.38,3.47,3.55,
3.66,3.78,3.92, 4.08,4.24,4.40,4.66,4.82,4.95,
5.11,5.34,5.59,5.80,5.99,6.16,6.36, 6.56,6.79,
7.01,7.13,7.34,7.56,7.76,7.98,8.21,8.46,8.70,
8.98,9.20, 9.44,9.67,9.93,10.21,10.38,10.61,
10.77,11.00,11.25,11.50,11.71,11.95,12.12,
12.34,12.58,12.85,13.05,13.31,13.47,13.60,
13.82,13.98,14.11,14.23,14.30,14.46,14.70,
14.92,15.07,15.10,15.35,15.50,15.69,15.88,
16.02,16.11,16.21,16.37,16.59,16.80,16.87,
17.02,17.23,17.44,17.59,17.72,17.86,17.98,
18.28,18.67,19.37,19.88,20.39,20.90,21.60,
21.92, 22.43]

[NGC 2403, respective velocities, in km/s] = 
[1.94,15.79,22.31,32.22,39.72,46.30,53.38,59.95,
64.03,67.36,71.44,75.65,77.92,79.58,80.51,82.13,
84.21,85.69,87.41,89.95,91.71,94.07,95.05,97.13,
96.71,98.61,98.56,98.10,99.81,101.02,102.78,
104.63, 106.20,107.92,109.81,110.97,112.73,
114.63,116.20,117.73,118.94, 120.05,121.25,
122.59,124.21,125.74,126.11,126.06,125.46,126.39,
127.27,128.61,127.18,126.94,126.90,126.02,
126.71,127.31,128.01, 127.78,129.12,129.63,
129.03,128.33,127.73,128.10,127.73,127.55, 
127.59,128.24,129.03,130.42,130.74,131.11,
130.88,131.57,132.13, 133.15,132.87,133.43,
133.70,134.72,134.81,135.97,137.59,139.12,
141.39,142.36,141.53,142.31,141.02,140.09,
138.89,138.56,141.06, 140.28,139.86,140.65,
140.83,138.94,136.62,135.60,136.30,135.42,
134.44,135.56,133.80,134.49,135.00,136.90,
138.15,137.22,138.61, 134.58,135.79,135.31,
133.47,133.26,134.08,135.51,137.96,139.18]
\end{verbatim}

	The data set for NGC 2903 was \cite{db}:

\begin{verbatim}
[NGC 2903, galactocentric distances, in kpc] =
[0.31,0.63,0.94,1.25,1.57,1.88,2.19,2.51,2.82,
3.12,3.43,3.75,4.06,4.37,4.69,5.00,5.31,5.63,
5.94,6.25,6.57,6.88,7.19,7.51,7.82,8.13,8.45,
8.76,9.07,9.39,9.70,10.01,10.31,10.63,10.94,
11.25,11.57,11.88,12.19,12.49,12.82,13.13,
13.45,13.76,14.07,14.39,14.70,15.01,15.33,
15.64,15.96,16.27,16.58,16.90,17.21,17.52,
17.82,18.13,18.45,18.76,19.07,19.39,19.70,
20.01,20.33,20.64,20.96,21.27,21.58,21.90,
22.21,22.52,22.84,23.15,23.46,23.78,24.09,
24.40,24.72,25.03,25.33,25.64,25.96,26.27,
26.58,26.90,27.21,27.52,27.84,28.15,28.46,
28.78,29.09,29.40,29.72,30.03,30.34]

[NGC 2903, respective velocities, in km/s] =
[20.22, 25.87, 28.26, 38.15, 65.98, 98.80, 
131.20, 149.67, 167.17, 183.15, 191.30, 
201.85, 206.52, 210.98, 213.48, 214.02, 
213.37, 213.15, 214.02, 215.22, 215.11, 
213.80, 212.28, 210.87, 207.61, 204.67, 
200.65, 198.91, 200.43, 200.11, 200.43, 
201.74, 203.91, 206.63, 205.98, 205.00, 
203.91, 203.26, 202.83, 202.17, 201.52, 
200.76, 200.22, 199.89, 199.89, 199.67, 
199.67, 199.67, 199.67, 199.67, 199.67, 
199.35, 198.91, 198.37, 197.07, 195.43, 
194.13, 192.83, 191.85, 191.09, 190.00, 
190.00, 189.89, 188.70, 187.83, 188.15, 
188.26, 188.48, 189.35, 190.11, 190.11, 
189.89, 189.78, 189.67, 189.67, 189.57, 
189.57, 189.46, 188.91, 187.07, 187.07, 
186.85, 186.74, 185.87, 184.46, 184.89, 
184.89, 184.89, 185.87, 186.85, 187.07, 
187.72, 185.22, 186.41, 188.48, 185.65, 179.35]
\end{verbatim}

The data set for NGC 5055 was \cite{db}:

\begin{verbatim}
[NGC 5055, galactocentric distances, in kpc] =
[ .27, .51, .75, .99, 1.25, 1.47, 1.71, 1.98, 
2.19, 2.46, 2.67, 2.94, 3.30, 3.69, 3.90, 4.17, 
4.39, 4.67, 4.87, 5.11, 5.37, 3.20, 5.86, 6.10, 
6.36, 6.58, 6.82, 7.08, 7.33, 7.54, 8.05, 8.53, 
9.01, 9.49, 10.00, 10.51, 10.96, 11.47, 11.93, 
12.19, 12.67, 13.16, 13.66, 14.63, 14.39, 14.87, 
15.13, 15.37, 15.86, 16.10, 16.34, 16.58, 16.82, 
17.08, 17.30, 17.57, 18.02, 18.29, 18.77, 19.25, 
19.73, 20.22, 20.46, 20.72, 21.45, 21.95, 22.41, 
22.94, 23.40, 23.86, 24.51, 24.87, 25.35, 25.86, 
26.10, 26.34, 26.58, 27.06, 27.54, 28.00, 28.51, 
29.01, 29.49, 29.98, 30.48, 30.94, 31.20, 31.42, 
31.95, 32.41, 32.92, 33.16, 33.66, 34.12, 34.60, 
34.87, 35.08, 35.59, 36.07, 36.80, 37.06, 37.30, 
37.52, 38.02, 38.51, 38.75, 38.99, 39.47, 39.98, 
40.46, 40.96, 41.42, 41.93, 42.39, 42.92, 43.40, 
43.86, 44.36, 44.84, 45.57, 46.07, 46.55, 47.04, 
47.52, 48.00, 48.53]

[NGC 5055, respective velocities, in km/s] =
[ 60.22, 123.80, 148.15, 160.22, 165.22, 173.48,
179.02, 182.72, 188.91, 194.46, 198.80, 202.93, 
205.54, 202.07, 197.93, 195.22, 193.59, 193.48, 
195.22, 197.93, 201.09, 203.26, 204.46, 206.20, 
206.74, 206.41, 206.20, 206.52, 206.09, 206.63, 
207.61, 208.37, 207.83, 207.28, 207.39, 207.93, 
209.02, 208.91, 210.22, 210.43, 210.43, 209.02, 
207.50, 208.37, 208.48, 208.04, 208.37, 206.41, 
203.80, 201.30, 199.13, 198.15, 197.50, 198.59, 
199.78, 201.09, 202.83, 203.15, 201.85, 199.67, 
196.09, 194.35, 193.59, 194.35, 195.43, 196.96, 
197.39, 196.85, 199.67, 200.76, 198.91, 200.11, 
198.70, 201.09, 203.48, 204.13, 203.15, 200.11, 
196.20, 192.17, 190.54, 190.98, 191.96, 193.37, 
192.39, 191.74, 189.89, 187.07, 186.52, 186.30, 
190.00, 193.26, 193.26, 191.52, 190.65, 188.80, 
186.20, 187.50, 189.24, 187.50, 185.11, 179.24, 
174.46, 170.65, 175.87, 177.93, 181.20, 182.83, 
183.70, 185.98, 188.26, 190.98, 192.50, 190.87, 
188.26, 186.52, 175.87, 165.11, 158.70, 162.72, 
157.93, 160.98, 165.22, 165.65, 166.52, 169.67]
\end{verbatim}

The data set for the Milky Way was \cite{sho}:

\begin{verbatim}
[Milky Way, galactocentric distances, in kpc] =  
[.094, .122, .141, .226, .311, .414, .518, .565, 
.696, .894, .988, 1.2, 1.3, 1.4, 1.6, 1.8, 2.0, 
2.1, 2.2, 2.3, 2.5, 2.7, 2.8, 2.9, 3.1, 3.3, 3.4, 
3.6, 3.8, 4.0, 4.1, 4.2, 4.3, 4.4, 4.5, 4.6, 4.7, 
4.8, 4.9, 5.0, 5.1, 5.2, 5.3, 5.4, 5.5, 5.6, 5.8, 
6.0, 6.1, 6.3, 6.5, 6.7, 6.8, 7.0, 7.2, 7.4, 7.6, 
7.8, 8.0, 8.1, 8.2, 8.3, 8.4, 8.6, 8.8, 8.9, 9.1, 
9.2, 9.4, 9.5, 9.6, 9.8, 9.9, 10.2, 10.3, 10.4, 
10.6, 10.7, 10.9, 11.0, 11.1, 11.3, 11.5, 11.7, 
11.8, 12.0, 12.3, 12.5, 12.6, 12.7, 12.9, 13.0, 
13.1, 13.4, 13.6, 13.8, 14.4, 14.6, 15.1]

[Milky Way, respective velocities, in km/s] =
[201.765, 218.694, 232.647, 245.435, 248.224, 
251.565, 248.006, 239.588, 229.259, 222.465, 
217.029, 216.288, 214.282, 208.588, 201.800, 
201.388, 199.488, 201.064, 195.427, 190.574, 
191.103, 189.794, 192.241, 192.406, 193.146, 
201.441, 197.688, 197.198, 201.041, 207.571, 
214.147, 213.212, 207.024, 212.728, 215.371, 
201.588, 200.557, 206.906, 210.171, 212.264, 
211.776, 207.165, 211.124, 207.471, 206.465, 
209.976, 209.029, 214.141, 218.609, 218.229, 
207.647, 208.176, 210.106, 208.424, 206.000, 
206.637, 203.481, 204.430, 200.000, 201.386, 
193.441, 189.373, 192.557, 196.398, 179.804, 
192.245, 193.000, 177.061, 173.333, 190.702, 
167.460, 184.670, 187.174, 197.143, 183.810, 
167.143, 166.190, 170.272, 177.619, 196.190, 
187.619, 183.382, 189.632, 191.264, 185.754, 
181.021, 193.180, 186.377, 190.476, 197.842, 
200.000, 211.352, 228.571, 204.000, 189.524, 
177.143, 209.000, 131.429, 239.700]
\end{verbatim}

\section{Coefficients of the expansions}\label{ap:4}

We obtained the rotation curves from the expansion in equation (\ref{eq:a11}). Below we present the coefficients of this equation found applying the fitting procedure to the data listed in appendix \ref{ap:3}.

	Coefficients that we calculated using maple in the study of  NGC 2403:

The $k_n$ parameters for NGC 2403:
\begin{verbatim}
[0.1068811359, 0.2453368049, 0.3846101294, 
0.5240681973, 0.6635963426, 0.8031583986, 
0.9427394057, 1.082332068, 1.221932406, 
1.361538065, 1.501147565, 1.640759927, 
1.780374478]
\end{verbatim}

The coefficients $C_n$ for NGC 2403:
\begin{verbatim}
[-0.007921808141,-0.0005506380009,
-0.0004107285579,-0.0001416047879, 
-0.0001247079410,-0.000003841649130,
-0.00004885012148,-0.00001683800647, 
-0.00002487709404,-2.343858445 10^(-8), 
-0.00003700513188, 0.000001930358895, 
-0.00004656681353]
\end{verbatim}

Coefficients that we calculated using maple in the study of  NGC 2903:
The $k_n$ parameters for NGC 2903:
\begin{verbatim}
[0.07858907052, 0.1803947095, 
0.2828015658, 0.3853442627, 0.4879384872, 
0.5905576460, 0.6931907395]
\end{verbatim}

The coefficients Cn for NGC 2903: 
\begin{verbatim}
[-0.01530101756, -0.001702251876, 
-0.001093545710, -0.0003322637691, 
-0.0002932484585, -0.0001243144261, 
-0.00006261309948]
\end{verbatim}

Coefficients that we calculated using maple in the study of  NGC 5055:
The $k_n$ parameters for NGC 5055:
\begin{verbatim}
[0.04948200737, 0.1135818541, 0.1780602451, 
0.2426241655, 0.3072205291, 0.3718325920, 
0.4364534287, 0.5010796612, 0.5657094472, 
0.6303416970, 0.6949757246, 0.7596110773, 
0.8242474437, 0.8888846032, 0.9535223946, 
1.018160698, 1.082799421, 1.147438493, 
1.212077859, 1.276717473, 1.341357301, 
1.405997312, 1.470637482, 1.535277791, 
1.599918223, 1.664558763, 1.729199399, 
1.793840120, 1.858480919, 1.923121786, 
1.987762715, 2.052403701, 2.117044739, 
2.181685821, 2.246326947, 2.310968113, 
2.375609315]
\end{verbatim}

The coefficients Cn for NGC 5055: 
\begin{verbatim}
[-0.02364170264,-0.003062378368,
-0.001529398084,-0.0007544629681, 
-0.0004622872640,-0.0003383465446,
-0.0001185939475,-0.0001872887342, 
-0.0001070291160,-0.0001398569473,
-0.00006206609205,-0.00002894284164, 
-0.0001352079235,+0.00003397666970,
-0.0001022605028,+0.00001757782189, 
-0.00007603025313,-0.00001683145888,
-0.00002683152942,-0.00001692287087, 
-0.00001810184718,-0.000007174947439,
-0.00002133565647, -0.00002204377444, 
+0.000005757757480,-0.000001880154179, 
-0.000005769757992,-0.00002583865542, 
+0.00001112592202, -0.000009783967047,
-0.000009787247767, +0.00001670245632, 
-0.00001880337343,-0.000005612283845,
-0.00001008754053, -0.00001203562779,
-0.00002228534325]
\end{verbatim}

Coefficients that we calculated using maple in the study of  the Milky Way:

The $k_n$ parameters for the Milky Way: 
\begin{verbatim}
[0.1582122078, 0.3631630335, 0.5693242048, 
0.7757588447, 0.9822972177, 1.188885787, 
1.395502410, 1.602136285, 1.808781522, 
2.015434636, 2.222093435, 2.428756470, 
2.635422747, 2.842091560, 3.048762393, 
3.255434862,  3.462108673, 3.668783602, 
3.875459469, 4.082136131, 4.288813473, 
4.495491403, 4.702169842, 4.908848726, 
5.115528002, 5.322207623, 5.528887551, 
5.735567752, 5.942248200]
\end{verbatim}

The coefficients $C_n$ for the Milky Way: 
\begin{verbatim}
[-0.007697600875, -0.0008050222929, 
-0.0006454430625, -0.00008960072403, 
-0.0001409159428, -0.0001215910888, 
-0.00007759716357, -0.00006105400402, 
-0.00006353141349, -0.00003911832613, 
-0.00005552511805, -0.000007937031005, 
-0.00005138018826, 0.00001212383799, 
-0.00003946152395, 0.00001509900263, 
-0.00004951153900, 0.000005909808077, 
-0.00002483567297, -0.00001052274983, 
-0.00001230139145, -0.000005792751710, 
-0.00002321451846, 0.00001099990558, 
-0.00003005489058, -0.000001929859973, 
-0.00002397747177, 0.000004644731306, 
-0.00004513811946]
\end{verbatim}

\end{document}